# Repetitive users network emerges from multiple rumor cascades

Li Daqing[1,2*], Gao Jiali[1], Zhao Jichang[3], Zhao Zilong[1], Orr Levy[4], Shlomo Havlin[4]

1. School of Reliability and Systems Engineering, Beihang University, Beijing 100191, China
2. Science and Technology on Reliability and Environmental Engineering Laboratory, Beijing 100191, China
3. School of Economics and Management, Beihang University, Beijing 100191, China
4. Department of Physics, Bar Ilan University, Ramat Gan 5290002, Israel

**Rumor spreading on online social media is presenting a significant threat to society of post-truth epoch. Extensive efforts have been devoted to rumor identification and debunking, assuming that a specific rumor propagation is a single event network and neglecting possible interdependence between different rumor cascades. Here we study the collective propagation of multiple rumors, and surprisingly find a network of users that repeatedly participate in different rumor cascades. Though these repetitive users demonstrate minor difference at the level of single propagation network, they are found to form a significantly more intensive collaboration network from multiple rumor cascades compared to news propagation. The clique-like cluster formed by repetitive rumor spreaders can serve as a high quality feature for rumor identification and blocking targets for rumor prevention. Our findings can provide a better understanding of rumor spread by viewing multiple rumor propagations as one interacting rumor ecosystem, and suggest novel methods for distinguishing and mitigation based on rumor spreading history.**

Rumors, mainly transmitted along the history by word of mouth, and have been involved in a "chain of subjects" who passed a story from "mouth to ear" without possibility to verify it[1]. Recently, the blooming of online social media like Twitter and its variant Weibo of China, has profoundly reshaped the manner of information communication[2-6], especially manifested in rumors[7-10]. For example, in October 2008, a rumor that Apple's CEO Steve Jobs suffered a major heart attack has been circulated. Although this rumor has been proven false, its rapid spreading and initial adoption as a fact by investors had substantial impact, resulting in a loss of $9 billion market value[11]. Rumor propagation can undermine the basic value of modern society networked by the mutual belief. According to the World Economic Forum, dis-intermediated circumstances unfortunately facilitate the spread of conspiracy theories, misinformation and rumors, which present one of the primary threats in our modern society[12-13].

With the recent explosion of new media channels, such as Facebook and Twitter, the identification and prevention of rumors became much more difficult. Thus, understanding the mechanisms of rumor spreading in online social media has attracted much efforts, developing various approaches such as model-based simulations including the well-presented Daley Kendall (DK) model[14]. Further studies consider various propagation characteristics on networks of diverse topologies[15-19]. Epidemic models have been also pervasively employed in modeling rumor cascades in terms of constructing estimators for rumor sources or understanding the impact of network structures[20-22]. Later, with the pervasive permeation of social media and the continuous accumulation of rumor-related propagation footprints, data-driven solutions become possible and popular[23-25]. Recent studies have demonstrated the existence of echo chamber in rumor spread, i.e., homogeneous clusters that users interact mainly within similar kind of content[26-27]. Polarized users have been found significantly different in consuming scientific-like or conspiracy-like messages[28]. To combat these rumor cascades, different methods have been proposed. For example, measures of information credibility have been developed for filtering out rumors[29], where both content and prorogation network have been used as features in rumor detection[30-32]. Meanwhile, based on digital traces collected from Facebook and Snopes, recent studies found that a rumor might cascade for weeks or months, and then become popular again through an external jolt[8]. Rumor prevention can be modeled as the minimization of 'bad' influence through identifying a subset of individuals that can be convinced to adopt the 'good'[33]. Unfortunately, it is found that debunking efforts seem ineffective in many rumor propagations[34], while some debunking efforts may even reinforce the strength of rumor propagation.

In the study of rumor dynamics, most previous efforts roughly simplified the rumor propagation into single processes, neglecting the fact that different rumors might be disseminated interdependently by the same group of users, over the online social network. We argue here that a more systematic and realistic view could be established through integrating the different single rumor's cascades into collective multiple interactions. Hence in the present study, based on empirical dataset collected from Weibo, one of the most trended social media in China, the interactions between different rumor retweeting networks (or news retweeting networks), due to repetitive spreaders, are thoroughly investigated. It is found that repetitive users participating in the rumor propagation show the trace of stronger collaboration than repetitive users in the propagation of authentic messages, which essentially helps the penetration of rumors in social media. These repetitive spreaders evolve into clique-like clusters forming a multilayer network, offering a novel proxy for rumor identification and blocking. Our findings based on analyzing collective social dynamics provide a broader perspective of rumor spreading understanding, which may enhance existing knowledge that emphasis semantic or propagation features of single cascades.

**Results**

**Repetitive spreaders in rumor propagation.** For each tweet in social media, either rumor or news, its propagation trace could be modeled as a retweeting network. A repetitive user is defined as a node participating in more than one retweeting network, as demonstrated in Fig 1(a). Intuitively, activeness (defined as *R*, see Methods for more details) of repetitive spreaders measures the frequency that the corresponding node attends in multiple propagations, in all rumor or news networks. Those users with high activeness (defined as key nodes) in different networks demonstrate greater vitality in the social media. Then, as Fig 1(b) shows, users attend in rumor or news networks are inclined to have many occurrences across different propagation events, suggesting the important role of these overlap nodes. Both distributions seem to follow an approximated power law function with similar exponents. Note that nodes with extremely high $R$ in news networks can be official media accounts that appear in many news propagation networks. These official media channels post much news and thus possess high activeness. However, as seen in Fig 1(c), connected pairs that represent retweeting relationships between two users, are found to have zero correlation of activeness for news propagation (see Methods and Fig. S1 for more details), while a positive correlation can be found mainly for rumors. In contrast to news, in rumors, nodes with higher $R$ tend to link (with retweet) nodes that also possess high value of $R$. This suggests the possibility of collaboration between key nodes in rumors in contrast to real news. Thus, in both official and non-official news propagation, users behave more independently, compared to rumors. Note that the correlations of $R$ values between two ends of edge in propagation network are further exploited in Fig. S1.

**Propagation properties of highly repetitive nodes (key nodes).** While a group of users are found to show unexpected high activeness and appear repeatedly in multiple social information propagations, their structural properties might help to better understand their role in the information spreading. Here we define key nodes ($k$ nodes) as repetitive nodes with high activeness (top 0.1% of highest $R$), which are marked in Fig 1b. It is seen in Fig 2a that in contrast to news, key nodes in rumors appear typically at larger distance from creators of propagation than distance one for news networks (see Fig. S3 for examples). Their high presence is mostly at layer three, while the ordinary (non-key) nodes, see inset of Fig 2a, appear mostly in the first layer both in rumors and news. While the key nodes in layer one of news are mostly followers of news publisher, the concurrence of key nodes in layer three of rumor propagation network suggests possible coordination between these nodes. These key nodes also possess longer posting time $\Delta t$ from the creating time of networks than those in news networks, as seen in Fig 2b. People tend to believe the rumors when it is made familiar by repetition, known as the continued-influence effect (CIE) [35-37]. This suggests that key nodes usually participating in rumor propagations at later stages, possibly enhance the propagation intensity through repetition, behaving more like intensifiers rather than broadcasters in the news propagations.

However, as shown in Fig. 2c, minor differences have been found in degree properties between key nodes of rumors and news. Interestingly, this similarity between rumor and news is broken in Fig. 2d (see Fig. S2 for more) when plotting the averaged out-degree of nodes as a function of $R$ value. This result shows that the repetitive users in rumor cascades seems to have a constant averaged out-degree close to one for all $R$, while key repetitive users in official news show two distinct tendencies. One group of key users with large activeness in official news has large out degree, illustrating their strong broadcasting impact. Another group of key users with small degree are obviously located on the edge of propagation network. When comparing results shown in Fig. 2c and 2d, we find that the structural behaviors of key nodes, like degree, in a single propagation network may appear similar between rumor and fact (Fig. 2c). However, these two groups of key users become significantly distinguishable (Fig. 2d) when we consider their repetitive behaviors in multiple cascades.

**Emergence of a highly collaborative network behind rumors.** Besides presence in single propagation, these key nodes may actually interact and collaborate in multiple events of rumor cascades. To understand this, we construct a 'co-author' network (*C* network) to study the possible collaborative relations between these key nodes, which emerge during the evolution of multiple propagation networks. Nodes in this network are key nodes (or k nodes), links are added if two key nodes appear in the same rumor propagation. This kind of collaborative network is common to analyze in social networks such as scientific collaboration networks [38], and can signify how these repetitive users collaborate. The first question we ask here is how this collaboration network emerges with time. As demonstrated in Fig 3a, we rank the two sets of propagation events (rumor or news) according to their creating time, and build *C* networks by the following rules: first we include all of key nodes into the *C* network; if two key nodes appear in the same original propagation network, we add an edge between these two nodes accordingly in the *C* network. This consideration is based on possible crowdturfing behavior that malicious users may be organized to enhance the rumor propagation through disguising themselves as weakly coupled ordinary users [39]. Surprisingly, we find in Fig 3b that all key repetitive users in rumors can be identified in the very early stage of the whole period. This suggests that we need only a relatively small sampling datasets of rumor propagation to identify most of these highly suspected repetitive users. This finding is independent of the starting point of collection process, as shown in the inset of Fig 3b and Fig. S4d. Compared with news, this fast integration of repetitive users in rumors is made of non-uniform bursts of increase, showing the correlated appearance of key users.

The key repetitive users not only show fast integration from multi-cascades of rumors, but also have more concurrence probably due to higher coordination among them in a single propagation event. As shown in Fig. 3(c), connections between key repetitive users increase much faster in rumor than that in fact, suggesting that on average each key user will have more partners of other key users in a given propagation network. As a result, it is seen in Fig 3(d) that the distance between these key nodes in the rumor collaboration network is much smaller than that of news, demonstrating the trace of organized behavior due to intensive collaboration between $k$ users in rumor propagation. This is further demonstrated in Fig 3(e) that, while for news these highly collaborative users form gradually modular sub-structures [40], rumor key nodes are becoming a densely connected single group rapidly. For further properties of the $C$ networks see Fig. S4.

**Prediction and mitigation of rumor propagation based on key nodes.** Hundreds of thousands of different rumors have been found in online social networks [7]. Facing such big data of rumor cascades, it remains challenging how to generate useful information from these historical datasets for combating future rumors. Once we accumulate a fraction of rumor propagation data, it becomes possible to identify the key repetitive users from these interconnected propagation network and build their collaboration network, which will bring new features for distinguishing between rumors and news. To detect the influence of key nodes, we define $R_k$ as the relative size of the branches of the *k* nodes (see Methods) in a given propagation network and find $R_k$ can be useful feature to distinguish rumors from facts. It is shown in Fig 4(a) that the probability of a given propagation network to belong to rumors is increasing fast with increasing $R_k$. Probability of being a rumor approaches 1.0 for $R_k$ above 0.1. The distribution of $R_k$ can be seen in Fig. S5, showing clearly that key nodes in rumors have typically larger $R_k$, thus more influential. Moreover, with the evolution of *C* network, we calculate the giant component G (as shown in Fig. S4a) of their collaboration network as the number of propagation networks ($n_i$) increases. When we use $G/n_i$ for a given propagation network shown in Fig. 4b, it is found to be even a better distinction between rumors and news.

Once we have a method to identify the susceptible rumors based on key nodes or other features, one can use it to perform a real-time mitigation by blocking potential influential spreaders. It is generally difficult to identify initiators of rumors at early stages, however, blocking the intensifiers in the propagation process might effectively weaken the spreading of rumors and avoid further bursts of rumor spread. When we remove a certain fraction of nodes in a given propagation network, the resulting giant component of network can measure the effect of different mitigation strategies, as shown in Fig. 5. Here we compare the case where the nodes removal based on their $R$ values (from high $R$ to low $R$), with a random removal of the same fraction. It is seen in Fig. 5a that the removal of key nodes in rumor propagation networks generates a much larger damage compared to random removal. The damage made by removing key nodes in rumor is also much larger than that in news propagation (compare Figs. 5a and 5b). Note that this happens despite of the fact that the structural properties such as out-degree distribution are similar for key nodes of rumor and news cascades, as seen in Fig. 2c. Thus, the reason for the higher impact of key nodes in rumors compared to news, is mostly due to their significant differences in collective effect from the network of networks that can better mitigate the rumor cascades. This points towards those nodes playing an important role in multiple cascades, which can serve as an effective option in the comprehensive combat against rumor propagations.

**Discussion**

On one hand, rumors are low-cost and low-tech communication weapons that can be used by anyone to disrupt the efforts of businesses, civil affairs, nation or other credit systems. On the other hand, rumors, as a collective 'intelligence' process, have been evolving and improving themselves due to the game process [41] together with the debunking efforts [42]. With the increasing prevention abilities, rumors that survive, are becoming smarter than before, and are more difficult to distinguish, identify and eliminate. While this interaction and evolution is expected to become much more represented and resilient in the future ecology of rumors, our goal is to uncover its evolution features in order to enable effective debunking.

To achieve this goal, different technical approaches have been taken, including the development of denial platform from third party and tools based on machine learning methods. The denial platform of a third party usually relies on expensive yet slow manual recognition. Meanwhile, machine-learning based auto-recognition tools are insufficient against the lasting evolution of survival rumors, if they use a static combination of features. Rumor is analogous to the evolution of viruses, which requires continuous development of new drugs.

One possibility of eliminating such evolving rumors is to study deep into their "DNA", and identifying their fundamental mechanisms of reproduction. Here we find that among multiple rumor networks for different topics, there is a small group of special users (key nodes) showing trace of intensive collaboration in the different propagation networks, and thus, serving as a motor for rumor reproduction. This small group connecting multiple rumor cascades shows strong interactions inside the group (Fig. 3), while in single networks they behave similar to real news (Fig. 2c). Our results show that repetitive users could provide useful features for rumor identification and mitigation. Therein, this group of key repetitive users provides us an opportunity to recognize, distinguish and even prevent the rumor evolution and their collective propagation, by viewing the different rumors as a whole network of networks.

**Methods**

**Database introduction.** Focusing on the relationship of multiple rumor propagations, we begin by studying the different behaviors of repetitive spreaders in both rumor network and news network, in which nodes being users of Weibo and links being retweeting relationship. Rumors dataset containing 0.4k post were officially debunked by Weibo and news dataset containing 0.4k posts in total were posted by verified accounts of mainstream mass media. The retweeting traces of rumors and news were thoroughly collected through Weibo's open APIs.

**Definition of key nodes.** As shown in Fig 1a, one node might repetitively attend propagations of multiple message propagation networks either for rumors or news. These overlapped nodes are named repetitive users. To quantify the overlap level, i.e., the activeness of a repetitive user, denoted as $R$, is defined as the node appearance frequency

$$R = \frac{n}{N}, \qquad (1)$$

where $n$ is the total number of propagation networks that the user attends, $N$ is the total number of networks in the employed data set (for rumor $N$ is 407 networks and 426 for news networks). Moreover, we define the nodes with the top 0.1% highest activeness as key nodes, which present higher appearance frequency. In rumor propagation there are 302 key nodes in total and 477 key nodes in news propagation. We define the correlation of activeness between the connected nodes as follows:

$$corr = \frac{<R_1 * R_2> - <R_1> * <R_2>}{<R_1 * R_2>}, \qquad (2)$$

for each connected pair of nodes in multiple propagation networks, $R_1$ and $R_2$ are the $R$ of both ends of an edge.

**Propagation features.** As for rumor identification (see Fig. 4), in each network, $R_k$ is defined as

$$R_k = \frac{N_k + \sum N_{bs}^{ki}}{N_{net}}, \qquad (3)$$

where $N_k$ is the number of key nodes, $N_{bs}$ is branch size of key nodes in the network and $N_{net}$ is the total number of nodes that attend in the given network. It is worth noting that in social media, the branch size can measure a spreader's influence of contagion, because any message posted by the spreader will be firstly pushed to all their branched size and higher branch size implies more retweets in later propagations. Furthermore, probability of a random network belongs to rumor networks in each $R_k$ section ($P$) is defined as

$$\begin{cases} P_{ri} = \dfrac{a_i}{a_i + b_i} \\ a_i = \dfrac{n_{ri}}{N_r} \\ b_i = \dfrac{n_{fi}}{N_f} \end{cases}, \qquad (4)$$

where $n_{ri}$ is the number of rumor networks in $i_{th}$ $R_k$ section and $N_r$ is the total number of rumor networks, $n_{fi}$ is the number of news networks in $i_{th}$ section and $N_f$ is the total number of news networks.

**Acknowledgements**

We thank the support from National Natural Science Foundation of China (Grants No. 71771009). S. H. thanks the Israel Science Foundation, ONR, the Israel Ministry of Science and Technology (MOST) with the Italy Ministry of Foreign Affairs, BSF-NSF, MOST with the Japan Science and Technology Agency, the BIU Center for Research in Applied Cryptography and Cyber Security, and DTRA (Grant no. HDTRA-1-10-1- 0014) for financial support.    Z. J. was supported by NSFC (No. 71501005) and the National Key Research and Development Program of China(No.2016QY01W0205).

**Author contributions**

L. D., S. H., Z. J., Z. Z., O. L. and G. J. conceived and designed the research. G. J., Z. Z., O. L. and L. D. carried out the numerical simulations and data analysis. L. D., Z. J. and S. H. wrote the paper with contributions from all other authors.

**Additional information**

Correspondence and requests for materials should be addressed to L. D. (daqingl@buaa.edu.cn).

**Competing financial interests**

The authors declare no competing financial interests.

**Data Availability**

Due to privacy issues, we agree not to spread data publically according to the limitations of collection licensees. Meanwhile, our data are available from the corresponding author on reasonable request.

**Figures**

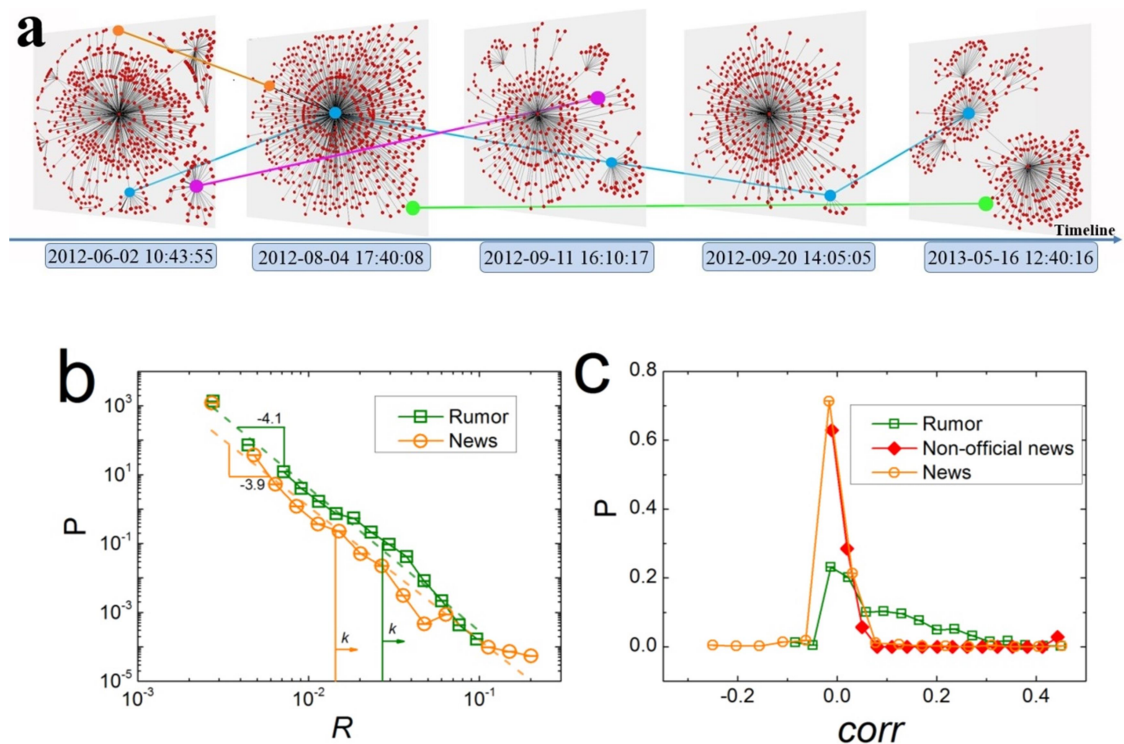

**Figure 1: Repetitive spreaders in rumor propagations.** (a) Demonstration of typical propagation networks of five rumors obtained from Weibo data. Repetitive spreaders are marked as non-red nodes in each network. We connect repetitive spreaders with lines and nodes of same color to demonstrate their repetitive occurrences. The creating time of each network is given in the figure. (b) PDF of $R$, the node appearance frequency in different rumors or news networks respectively. The label $k$ (key nodes) represents the considered range of most active nodes (top 0.1%) for both rumor networks and news networks. Here we use log binning with normalization. (c) Probability distribution of *corr*, *corr* is the activeness correlation between each connected pair in event propagation.

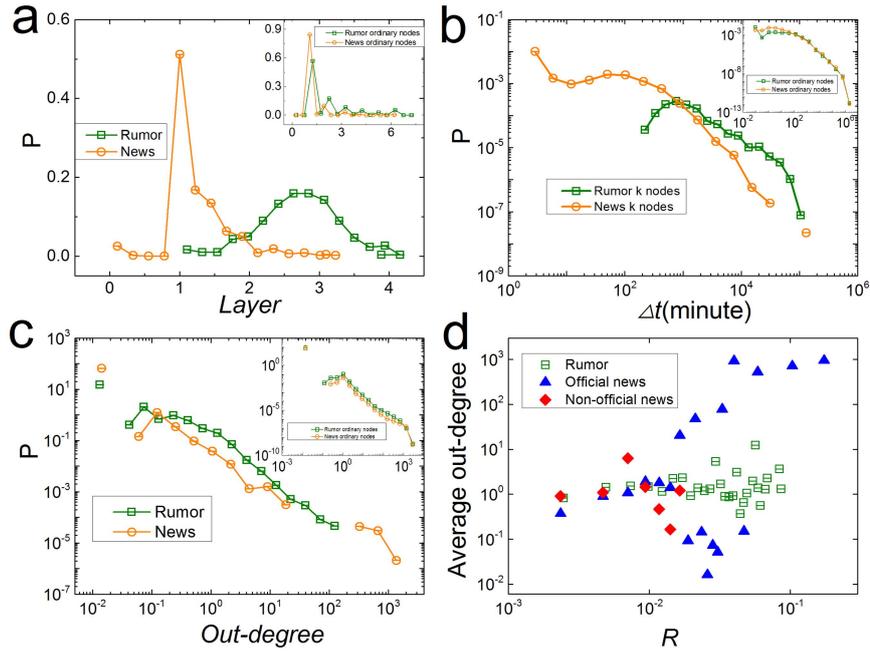

**Figure 2: Structural properties of highly repetitive nodes (key nodes) in propagation networks.** (a) The probability distribution of layer of key nodes. Layer represents the hop distance between retweeter and creator of a propagating network. For each key node, we first calculate its average layer in all networks that it attends and then obtain a probability distribution of the average layer of key nodes. We also present the layer distribution of ordinary nodes (all nodes except for key nodes) in the inset. (b) PDF of $\Delta t$ of key nodes, $\Delta t$ is defined as the interval between post-time of a key node and creating time of the corresponding network. Similarly, $\Delta t$ of each key node is also averaged over the multiple networks it attends. PDF of $\Delta t$ of ordinary nodes is also shown in the inset. (c) PDF of out-degree of key nodes in each network. (d) The averaged out-degree of nodes in the propagation networks as a function of $R$.

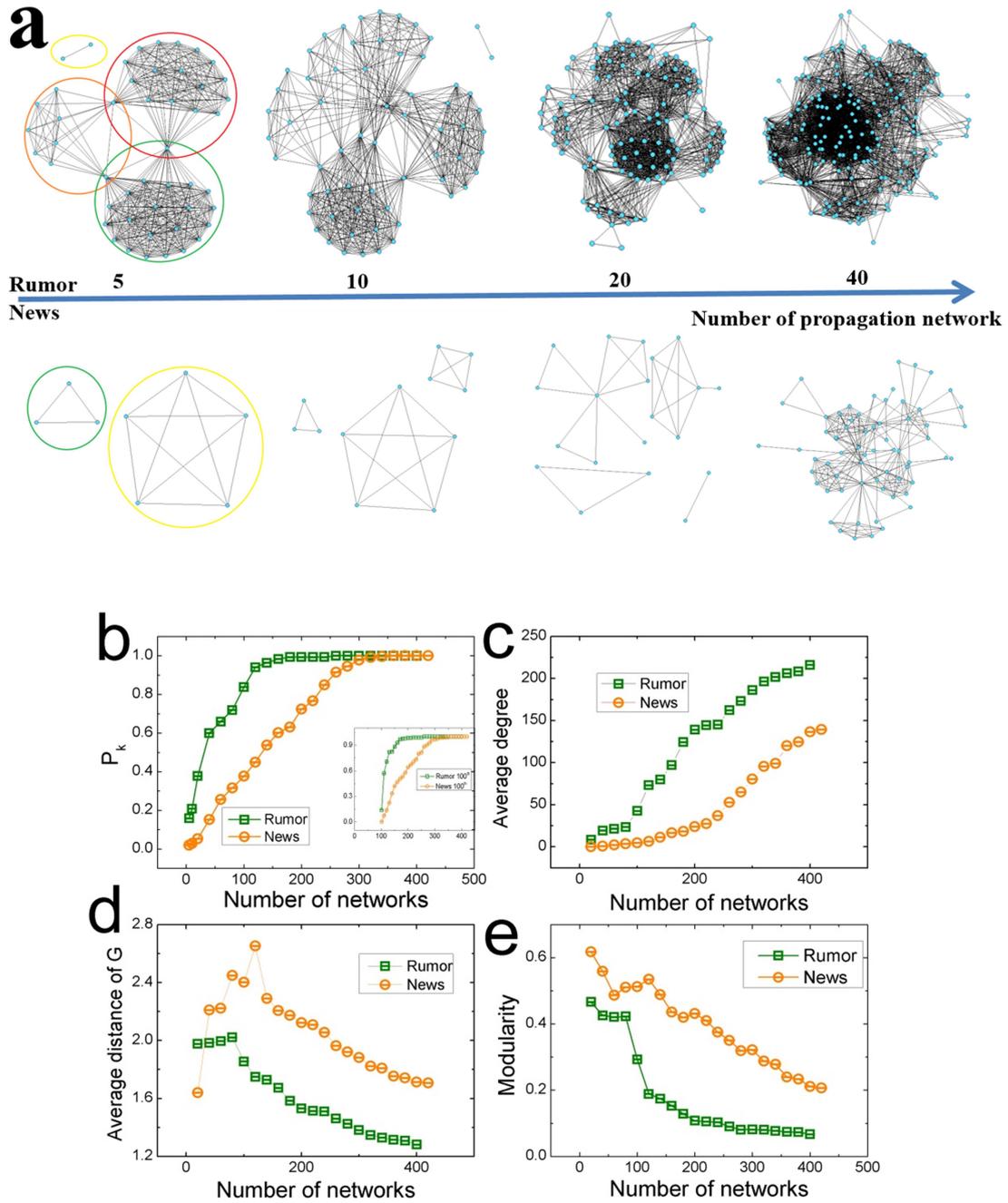

**Figure 3: Emergence of collaborative network.** (a) Evolution of *C* network with the increasing collaborations of key nodes as the number of propagation networks increases, ranked according to their creating time. Here we only show the large connected components in *C* network. (b) The fraction of distinct *k* nodes ($P_k$) grows as the number of sorted networks increases. We repeat the same process when starting from the 100[th] propagation network in inset figure. (c) Average degree of *C* network as the number of network increases. (d) Average distance between nodes in the giant connected component of *C* network decreases with the growing number of networks. (e) Modularity of *C* network decreases when the number of networks increases.

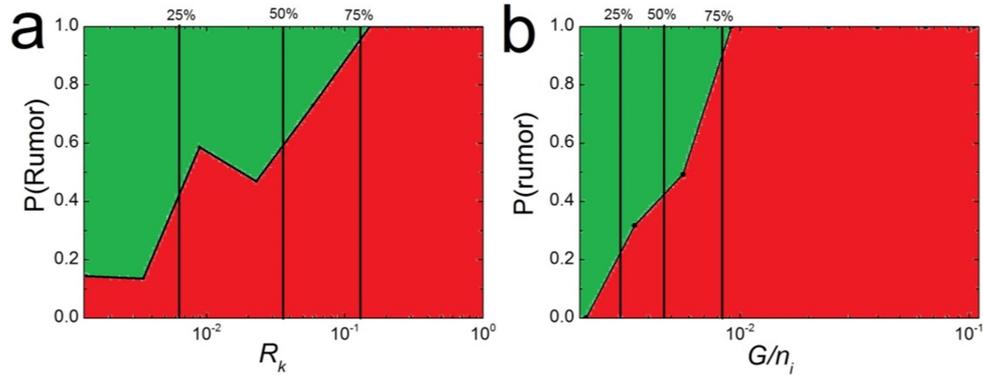

**Figure 4: Repetitive users as prediction feature.** (a) For each network, we calculate the fraction of $k$ nodes and their branch size in the network as the $R_k$ of the network. Here we only consider the non-official news and rumor. For each $R_k$ section, we calculate the probability (P) that a random network belongs to rumor network. Three vertical lines divide the whole dataset (rumor and non-official news) into 4 parts with equal number of networks sorted by $R_k$ values. For example, "25%" stands for having 25% of the networks in this area. (b) Size of collaborative networks. We calculate $G/n_i$ as the $C$ network grows when the number of networks increases, where $G$ is the giant component fraction of the $C$ network and $n_i$ is the number of added propagation networks. Here we only consider the non-official news and rumor in the $C$ network growing process.

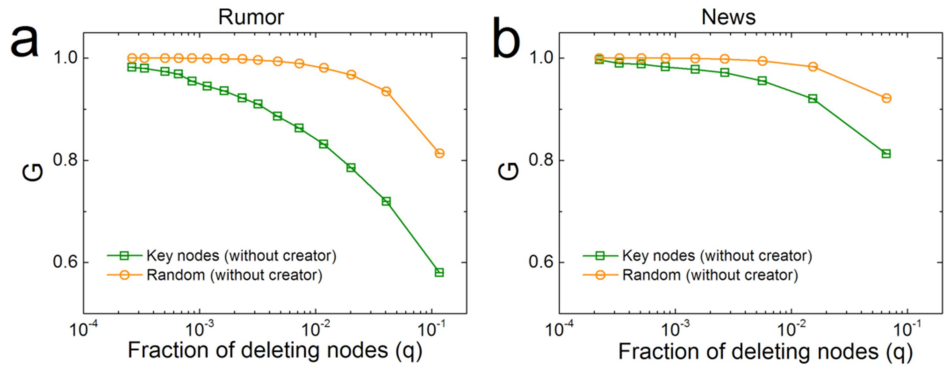

**Figure 5: Mitigation of rumor cascades using repetitive users.** Here we remove the nodes randomly, or based on their $R$ values (in a descending order). For each fraction, we delete the target nodes in each propagation network and calculate the giant component proportion (G) of the propagation network. Here we only consider and compare the propagation networks with repetitive users. Panel (a) and (b) are the G as a function of deleting fraction in rumor and news respectively.